\documentclass[]{WileyASNA-v1}

\usepackage[utf8]{inputenc}
\usepackage{calc}
\newlength{\okinalen}
\newcommand{\okina}{\hbox to.666\okinalen{\hss`\hss}}

\articletype{Article Type}%

\received{08 Dec 2019}
\revised{19 Feb 2020}
\accepted{27 Feb 2020}

\begin{document}

\title{Follow-up imaging observations of comet 2I/Borisov\protect\thanks{Based on observations obtained with telescopes of the University Observatory Jena, which is operated by the Astrophysical Institute of the Friedrich-Schiller-University.}}

\author[1]{M. Mugrauer*}

\author[1]{R. Bischoff}

\author[1]{W. Stenglein}

\author[1]{J. Trautmann}

\author[1]{B. Baghdasaryan}

\author[1]{S. Schlagenhauf}

\authormark{M. Mugrauer \textsc{et al}}

\address[1]{Astrophysikalisches Institut und Universit\"{a}ts-Sternwarte Jena}

\corres{M. Mugrauer, Astrophysikalisches Institut und Universit\"{a}ts-Sternwarte Jena, Schillerg\"{a}{\ss}chen 2, D-07745 Jena, Germany.\newline \email{markus@astro.uni-jena.de}}

\abstract{Follow-up imaging observations of the first detected interstellar comet 2I/Borisov are presented, which were carried out with the Schmidt-Teleskop-Kamera at the University Observatory Jena in 11 observing epochs in October and November 2019. The orbital solution of the comet, derived from the obtained astrometric measurements, confirms its highly eccentric ($e=3.3570 \pm 0.0006$) and\linebreak inclined ($i=44.0524 \pm 0.0004\,^\circ$) orbit, that proves the interstellar origin of the comet. According to our best-fitting orbital solution, comet 2I/Borisov reaches its perihelion on 2019 Dec 8 ($q=2.0066$\,au) and its closest encounter with Earth on 2019 Dec 28 ($\Delta_{\text{min}}=1.9368$\,au). The distance corrected brightness of the nucleus of the comet clearly exhibits a linear correlation with its phase angle. The slope of this correlation indicates the activity of the comet, which is also detected in deep imaging data, taken in 5 observing epochs, showing the coma and the tail of the comet. During our observing campaign the coma of comet 2I/Borisov exhibits on average a diameter of $(4.57\pm0.38) \cdot 10^4$\,km, and the length of the tail of the comet measures $(1.52\pm0.12) \cdot 10^5$\,km, assuming an anti-solar orientation.}



\maketitle

\section{Introduction}

The vast majority of small bodies, detected in the solar system, revolve around the Sun on low eccentric orbits mainly within the asteroid belt between the orbits of Mars and Jupiter, and in the Kuiper belt beyond the outer ice giant Neptune. In addition, also minor bodies on closed but highly eccentric orbits with orbital eccentricities $e>0.9$ were found in the solar system. Most of these objects are comets, among them 1P/Halley ($e\sim0.97$), whose periodic reappearance in the inner solar system was first noticed by Edmund Halley in 1705.

Minor bodies were also detected on open (i.e. parabolic or hyperbolic) orbits around the Sun, and almost all of them are comets. According to the JPL Small-Body Database Browser of the Jet Propulsion Laboratory, 230 such objects are known due to the end of November 2019, which exhibit significantly hyperbolic orbits ($e-3\sigma(e)\geq1$). These minor bodies mainly originate from the Oort cloud. Either they are scattered by close stellar flybys on hyperbolic orbits towards the Sun, or they are long periodic comets on highly eccentric orbits, whose eccentricity is then excited during close encounters with the giant planets in the solar system. According to \cite{laughlin2017}, for a planet to effectively act as gravitational perturber of small bodies, the ratio between the escape velocity from its surface and its average orbital velocity must significantly exceed unity. In general, all four giant planets of the solar system can cause gravitational perturbations on the orbits of minor bodies, which even can lead to their ejection from the solar system, with Jupiter and Neptune being the most important gravitational perturbers.

Indeed, the two most eccentric objects, known in the solar system before 2017, namely the comets C/1980~E1 ($e=1.058$), and C/1997~P2 ($e=1.028$) both went through close encounters (within a distance of 0.7\,au) with Jupiter several months prior to their perihelion passage, when their eccentricities were significantly excited. In the case of C/1980~E1, in addition to its close flyby at Jupiter, the comet also passed before Saturn within a distance of 3\,au. This extraordinary double encounter resulted in the strongest planetary perturbation, observed so far among minor objects on near parabolic orbits. After its close encounter with Jupiter C/1980~E1, which revolved the Sun originally on a highly eccentric orbit, was ejected from the solar system \citep{everhart1983}.

According to \cite{higuchi2019}, minor bodies on hyperbolic orbits with sufficiently small eccentricities but wide perihelion distances are most probably scattered objects from the Oort cloud, while minor objects with significantly larger eccentricities should be of interstellar origin in particular if close encounters with the giant planets can be excluded.

This holds for C/2017~U1, which was discovered in October 2017 (\href{https://www.minorplanetcenter.org/mpec/K17/K17UI1.html}{MPEC~2017-U181}) and was identified first as comet but later be reclassified as asteroid A/2017~U1, as no coma or tail was detected in very deep imaging data (\href{https://www.minorplanetcenter.org/mpec/K17/K17UI3.html}{MPEC~2017-U183}). This asteroid revolves around the Sun on a highly eccentric ($e\sim1.2$) and inclined ($i\sim123\,^{\circ}$) orbit with a perihelion distance of only $q\sim0.26$\,au. Due to its high eccentricity and because close encounters with giant planets can be excluded based on its orbital characteristics, the asteroid was classified as an interstellar object and named accordingly as 1I/2017~U1 (alias 1I/\okina Oumuamua), using the new small-body designation I, introduced by the international astronomical union for confirmed interstellar objects (\href{https://www.minorplanetcenter.org/mpec/K17/K17V17.html}{MPEC~2017-V17}).

Another intriguing object in this context is the comet C/2019~Q4 (Borisov), which was just recently detected at the end of August 2019 (\href{https://www.minorplanetcenter.org/mpec/K19/K19RA6.html}{MPEC~2019-R106}). Already an early orbit determination, based on observations taken until mid of September 2019, which are distributed along an arc with a length of only about $6.1^{\circ}$, could prove the high orbital eccentricity ($e\sim3.1$) of this comet, which revolves around the Sun on an orbit, which is highly inclined ($i\sim45\,^\circ$) to the ecliptic. According to this early orbital solution, the comet is expected to reach its perihelion mid of December 2019 at a solar distance of $q\sim1.9$\,au.

Because of its high orbital inclination and close perihelion distance significant gravitational perturbations from the giant planets in the solar system can be ruled out for this comet, whose high eccentricity then confirms its interstellar origin. Therefore, the Minor Planet Center assigned to it the designation 2I/Borisov (\href{https://www.minorplanetcenter.org/mpec/K19/K19S72.html}{MPEC~2019-S72}). In contrast to 1I/\okina Oumuamua comet 2I/Borisov was detected several months prior to its perihelion passage, which allows follow-up observations to study the properties of this intriguing object in its pre- but also post-perihelion orbital phase.

Therefore, we have started an imaging campaign of comet 2I/Borisov at the University Observatory Jena \citep{pfau1984} in order to study its astrometric and photometric properties, as well as its morphology during several weeks, while the comet remained observable from the location of the observatory.

In the following section of this paper we present the imaging observations of comet 2I/Borisov, which could be taken in the course of our imaging campaign, during several nights in October and November 2019 at the University Observatory Jena. The obtained astrometric data, as well as the derived orbital solution of the comet are presented in section 3. In the following section the photometry of the nucleus of comet 2I/Borisov is summarized for all observing epochs. The morphology of the comet, captured in deep imaging data, taken in the course of our observing campaign, is presented in section 5. Finally, all the results, derived from our imaging observations of comet 2I/Borisov, are discussed in the last section of this paper.

\section{Imaging Observations of Comet 2I/Borisov}

We have observed comet 2I/Borisov at the University Observatory Jena with the Schmidt-Teleskop-Kamera \citep[STK from hereon, ][]{mugrauer2010}, operated at the 90\,cm-reflector telescope of the observatory in its Schmidt-focus ($D=60$\,cm, $f/D=3$). The observations were carried out after an optimized focussing of the instrument in the Bessell R-band with detector integration times ($DIT$) of 60\,s, but mainly 120\,s. The exposure time was chosen according to the given sky brightness (mainly dependant on the angular distance between the comet and the moon, and the lunar phase), as well as on the proper motion of the comet. The details of the observations for all observing epochs are summarized in the observation log file, shown in Tab.\,\ref{log}\hspace{-1.5mm}.

Due to the position of the comet relative to the Sun, the observations always had to be carried out at a high airmass (2.4 to 1.6) but could be completed in most of the observing epochs before the beginning of the astronomical twilight. Because of the lower elevation of the target during the observations the atmospheric seeing at the position of the target on the sky ranged between 3.9 up to 4.9\,arcsec (median value of 4.3\,arcsec). The seeing was determined by measuring the full-width-at-half maximum ($FWHM$) of background stars (i.e. of point spread functions), detected in the STK images.

\begin{center}
\begin{table}[h]
\caption{Observation log file. The dates of all observing epochs are listed together with the used detector integration time $DIT$, the number $N$ of images taken in total, the conditions of the sky at the location of the target, the range of the airmass of the target during the observations, as well as the average $FWHM$ of background star, detected in the STK images.}\label{log}
\centering
\begin{tabular}{cccccc}
\hline
Date        & $DIT$   & $N$ & sky          & air-    & $FWHM$\\
2019        & $[$s$]$ &     & condition    & mass    & $[$arcsec$]$\\
\hline
Oct 09      & 120     & 22  & clear        & 2.4 - 1.9 & 4.8\\
Oct 22      & 60      & 37  & thin         & 2.1 - 1.7 & 3.9\\
Oct 26      & 120     & 6   & thin         & 1.9 - 1.8 & 4.2\\
Oct 28      & 120     & 22  & clear        & 2.1 - 1.7 & 4.6\\
Oct 30      & 120     & 10  & thin         & 1.8 - 1.6 & 3.9\\
Oct 31      & 120     & 17  & clear        & 2.0 - 1.8 & 3.9\\
Nov 01      & 120     & 34  & clear        & 2.0 - 1.6 & 4.8\\
Nov 10      & 120     & 17  & clear        & 2.1 - 1.8 & 4.0\\
Nov 14      & 120     & 6   & clear        & 2.1 - 2.0 & 4.8\\
Nov 19      & 120     & 11  & clear        & 2.2 - 2.1 & 4.3\\
Nov 23      & 120     & 20  & clear        & 2.1 - 1.9 & 4.9\\
\hline
\end{tabular}
\end{table}
\end{center}

In total 202 images could be taken with the STK during our observing campaign of comet 2I/Borisov, which results in 367\,min of total integration time on the target. In 179 of all STK images the astrometry and photometry of the comet could be measured. The remaining 23 STK images were rejected from the astrometric and photometric analysis of comet 2I/Borisov. In these images the target was either detected at too low signal-to-noise-ratio ($SNR$) due to some thicker cirrus clouds, which passed through the STK field of view during the observations, or bright background stars or satellite trails were located too close to the position of the comet, so that their fluxes significantly affect the astrometry and photometry of the comet.

All STK images were reduced with darks and domeflats, which were always taken during twilight at the beginning or end of each observing night. The data reduction was done with the \textsc{STK-pipeline}, a software, based on \textsc{ESO-eclipse} \citep{devillard2001}, which was developed at the Astrophysical Institute Jena for the immediate reduction and quality checks of STK data already during the night time operation at the University Observatory Jena.

The astrometrical calibration of all STK images was achieved with the software \href{http://www.astrometrica.at/}{\textsc{Astrometrica}}, using reference stars from the Gaia DR2 catalogue \citep{gaiadr2}, which is nowadays the best choice according to the number of reference stars with highly accurate astrometric positions available on the whole sky. Stars with G-band magnitudes in the range between 15 and 18\,mag were selected, to assure that a sufficiently large number of stars are detectable in each STK image with $SNR > 10$, needed to achieve an accurate astrometrical calibration of the images. The derived pixel scale of the STK detector for all observing epochs is $PS=1.5466\pm0.0004$\,mas/pixel \citep[consistent with the value obtained by][]{mugrauer2010} with a detector position angle of $DPA=-1.632\pm0.036\,^\circ$ (to be added to a measured position angle on the detector, to obtain the real position angle on the sky).

The astrometrical position of the comet was determined in all STK images by measuring the position of the barycenter of light of its nucleus. On average, the position of comet 2I/Borisov could be determined with an astrometric accuracy of about 0.3\,arcsec, which corresponds to about $1/5$ pixel of the STK detector.

According to \cite{baraffe1998} and \cite{jordi2010}, the R- and G-band magnitudes of stars best match with each other within a wide range of color. Therefore, as all observations were carried out in the Bessell R-band, for the photometric calibration of the STK images the G-band magnitudes of the selected reference stars from the Gaia DR2 catalogue were used, to derive the photometric zero-points. The photometric measurements of the nucleus of the comet were performed with aperture photometry. Thereby, the aperture radius was chosen to maximize the photometric accuracy, which was achieved for a radius equal to the average $FWHM$ of background stars, detected in the STK images. The photometry of the core of the comet is found to be stable during individual observing epochs on the reached level of photometric precision, which ranges between 0.05\,mag and 0.17\,mag, depending on the sky conditions during the observations. Under a clear sky $\Delta G < 0.1$\,mag is reached, while the photometric uncertainty exceeds 0.1\,mag for observations taken through thin cirrus clouds.

\section{Astrometry and Orbit Determination of comet 2I/Borisov}

In total, 179 astrometric measurements of comet 2I/Borisov, could be obtained from our STK imaging observations, which were carried out within a span of time of 45\,days in the course of our observing campaign at the University Observatory Jena. The astrometric positions of the comet are distributed along an arc with a length of about $33.1\,^\circ$ and are summarized in the appendix of this paper in Tab.\,\ref{astrometry}\hspace{-1.5mm}.

\begin{figure}[h]
\centerline{\includegraphics[width=0.49\textwidth]{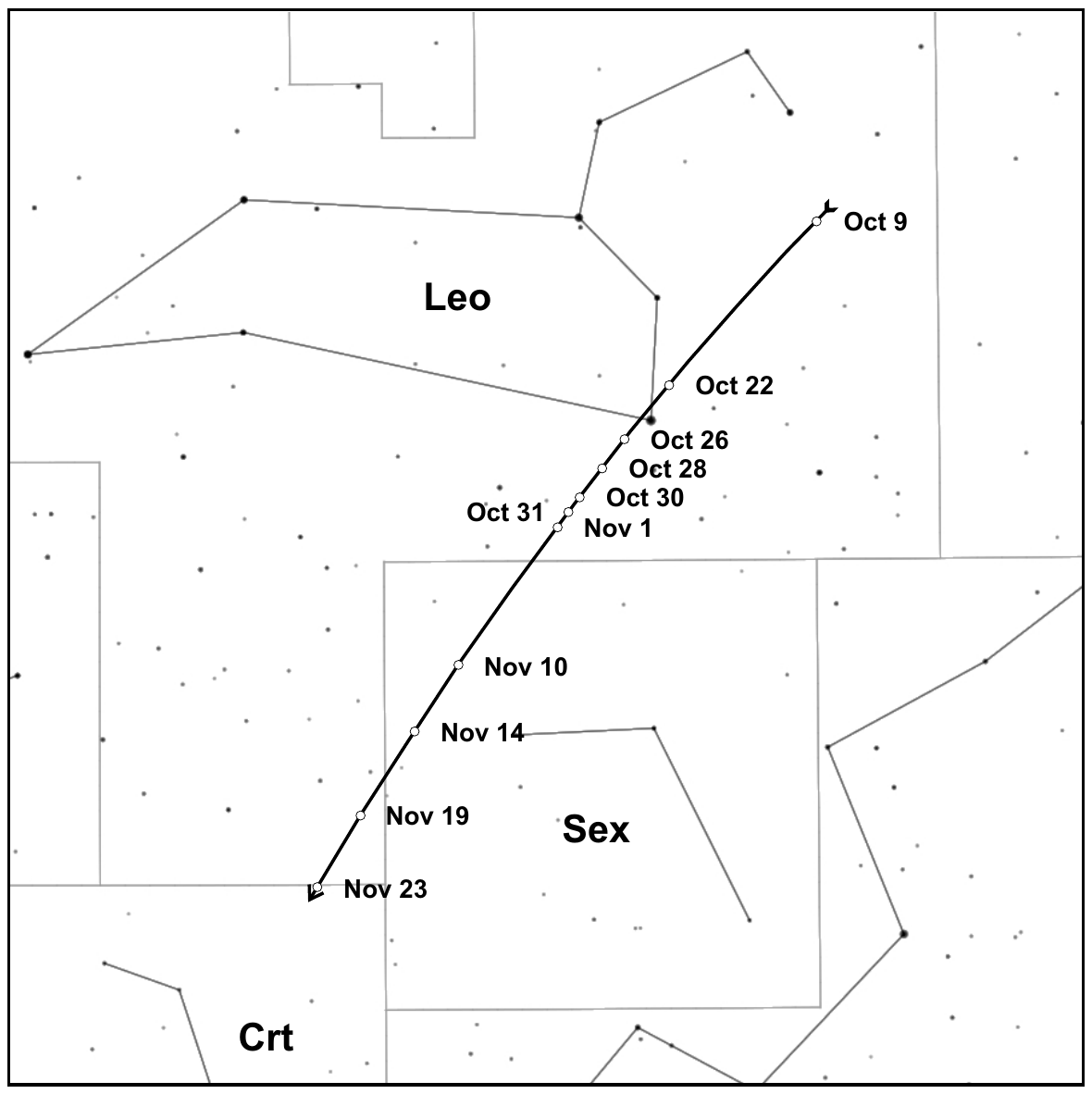}}
\caption{The path (black line) of comet 2I/Borisov on the sky during our imaging campaign, carried out at the University Observatory Jena, according to the best-fitting orbital solution of the comet, based on its STK astrometry from 11 observing epochs, which are illustrated with white circles. The path of the comet is shown for the range of time between 2019 Oct 8 and Nov 24, both at 0\,UT. During our observing campaign the comet moved southwards crossing the constellations Leo, Sextans, and eventually reached the constellation Crater. In this skychart East is to the left, and North is to the top.}\label{map}
\end{figure}

The orbit of comet 2I/Borisov was derived by least squares fitting of an orbital solution on the given STK astrometry of the comet, taking into account all planets, as well as the Moon, and Pluto as gravitational perturbers. The orbit determination was performed with the widely used software \href{https://www.projectpluto.com/find_orb.htm}{\textsc{find\_orb}}. The obtained best-fitting orbital solution exhibits a root-mean-square value (RMS) of 0.352\,arcsec, well consistent with the accuracy of the STK astrometry of the comet (0.33\,arcsec, on average), reached in the individual observing epochs. The position of the comet on the sky in all observing epochs together with the derived best-fitting orbital solution, are illustrated in the skychart, shown in Fig.\,\ref{map}\hspace{-1.5mm}, which is drawn with the software \href{https://www.ap-i.net/skychart/en/start}{\textsc{Cartes du Ciel V3.1}}.

The orbital elements of the derived best-fitting orbital solution of comet 2I/Borisov are listed with their uncertainties in Tab.\,\ref{elements}\hspace{-1.5mm}.

\begin{center}
\begin{table}[h!]
\caption{The orbital elements (for heliocentric ecliptic J\,2000) of the best-fitting ($RMS=0.352$\,arcsec) orbital solution of comet 2I/Borisov, based on 179 astrometric measurements, which are distributed along an arc with a length of about $33.1\,^\circ$, obtained from STK imaging data, taken between 2019 Oct 9 and Nov 23.}\label{elements}
\centering
\begin{tabular} {ll}
\hline
Epoch &  2019 Nov 1 (JD 2458788.5)\\
\hline
$T$               & $2019 \text{~Dec~} 8.555 \pm 0.002$\\
$q~[\text{au}]$    & $2.00657 \pm 0.00008$\\
$e$               & $3.3570 \pm 0.0006$\\
$i~[^\circ]$      & $44.0524 \pm 0.0004$\\
$\omega~[^\circ]$ & $209.1256 \pm 0.0007$\\
$\Omega~[^\circ]$ & $308.1484\pm0.0009$\\
\hline
\end{tabular}
\end{table}
\end{center}

The orbital properties of comet 2I/Borisov during the individual observing epochs, derived with our best-fitting orbital solution of the comet, are summarized in \linebreak Tab.\,\ref{orbprops}\hspace{-1.5mm}. During our imaging campaign the distance $\Delta$ of the comet to Earth decreased from 2.9\,au to 2.1\,au, while its solar distance $r$ only slightly decreased from 2.4\,au to 2.0\,au. During its approach to Earth and Sun, the phase angle of the comet increased from $19.7\,^\circ$ to $27.4\,^\circ$. The proper motion of the comet speeded up from 1.6\,arcsec/min to 2.1\,arcsec/min. Hence, it is not significantly detectable in individual STK images as it always remains smaller than the typical $FWHM$ of point-like sources, which were detected in these images.

\begin{center}
\begin{table}[h]
\caption{The orbital properties of comet 2I/Borisov, for all observing epochs, as derived from the best-fitting orbital solution of the comet, based on its STK astrometry. The distance $\Delta$ of the comet to Earth, its solar distance $r$, phase angle $\phi$, and proper motion $\mu$ are listed together with the Julian date of all observing epochs.}\label{orbprops}
\centering
\begin{tabular}{ccccc}
\hline
JD$-$2450000 & $\Delta~[\text{au}]$ & $r~[\text{au}]$ & $\phi~[^\circ]$ & $\mu~[\text{arcsec}/\text{min}]$\\
\hline
8765.63414 & 2.8530 & 2.4118 & 19.7  & 1.6\\
8778.64934 & 2.6011 & 2.2655 & 22.3  & 1.7\\
8782.65404 & 2.5290 & 2.2259 & 23.0  & 1.8\\
8784.65645 & 2.4940 & 2.2071 & 23.4  & 1.8\\
8786.68442 & 2.4594 & 2.1888 & 23.8  & 1.8\\
8787.65939 & 2.4431 & 2.1803 & 23.9  & 1.8\\
8788.67136 & 2.4263 & 2.1716 & 24.1  & 1.9\\
8797.67104 & 2.2875 & 2.1035 & 25.6  & 2.0\\
8801.66482 & 2.2323 & 2.0786 & 26.2  & 2.0\\
8806.66985 & 2.1691 & 2.0524 & 26.9  & 2.1\\
8810.70262 & 2.1233 & 2.0355 & 27.4  & 2.1\\
\hline
\end{tabular}
\end{table}
\end{center}

\section{Photometry of comet 2I/Borisov}

The apparent G-band magnitudes of the nucleus of comet 2I/Borisov, as measured in our STK images, are summarized in Tab.\,\ref{photometry}\hspace{-1.5mm} for all observing epochs of our imaging campaign. On average, the core of comet 2I/Borisov exhibits an apparent G-band magnitude of $G=17.34\pm0.07$\,mag and no significant changes of brightness were detected neither during individual observing epochs nor within the whole span of time, covered by our STK imaging campaign.

\begin{center} \begin{table}[h]
\caption{The G-band photometry of the nucleus of comet 2I/Borisov for all observing epochs of our STK imaging campaign.}\label{photometry}
\centering
\begin{tabular}{ccccc}
\hline

JD$-$2450000 & $G~[\text{mag}]$\\
\hline
8765.63414 & $17.48\pm0.09$\\
8778.64934 & $17.37\pm0.17$\\
8782.65404 & $17.37\pm0.10$\\
8784.65645 & $17.30\pm0.06$\\
8786.68442 & $17.30\pm0.15$\\
8787.65939 & $17.29\pm0.08$\\
8788.67136 & $17.37\pm0.07$\\
8797.67104 & $17.28\pm0.05$\\
8801.66482 & $17.36\pm0.06$\\
8806.66985 & $17.36\pm0.09$\\
8810.70262 & $17.22\pm0.09$\\
\hline
\end{tabular}
\end{table}
\end{center}

\section{Morphology of comet 2I/Borisov}

Beside its astrometry and photometry we also have studied the morphology of comet 2I/Borisov in deep imaging observations, which could be taken in 5 observing epochs (namely on Oct 9, 28, 31, Nov 1, and 10) in dark time (i.e. the moon was below the horizon, and the observations were carried out before the beginning of the astronomical twilight) during clear sky conditions. All STK images, taken in these observing epochs, in which the imaging of the comet was not affected by close stars or satellite trails are combined by averaging (22, 22, 17, 15, and 14 images, respectively) after being shifted to correct for the motion of the comet throughout the observations. The obtained deep STK images of comet 2I/Borisov are shown in Fig.\,\ref{pics}\hspace{-1.5mm}. In these images stars appear as trails, whose orientation indicates the direction of the motion of the comet during the observations. On average, a $SNR = 5$ detection limit of $G = 21.4 \pm 0.2$\,mag is reached in our deep STK images of comet 2I/Borisov, as determined with the Gaia DR2 photometry of the chosen reference stars. The extent of the comet is determined in all deep STK images by measuring the distribution of all pixels, located around the position of the comet, which exhibit a flux detection of $SNR \ge 5$. The coma and the tail of comet 2I/Borisov are both well resolved in all deep STK images, whose extents significantly exceed the $FWHM$ of stars, detected in these images. The obtained angular diameter of the coma and the apparent length of the tail of the comet, as detected in all deep STK images, are summarized in Tab.\,\ref{morphology}\hspace{-1.5mm}.

\begin{center}
\begin{table}[h]
\caption{The angular diameter of the coma $\varnothing_{\text{Coma}}$ and the apparent length of the tail $\Lambda_\text{Tail}$ of comet 2I/Borisov, measured in all deep STK images, taken in the course of our imaging campaign of the comet.}\label{morphology}
\centering
\begin{tabular}{cccc}
\hline
JD$-$2450000 & $\varnothing_{\text{Coma}}~[\text{arcsec}]$ & $\Lambda_\text{Tail}~[\text{arcsec}]$ &\\
\hline
8765.63414 & $23.2\pm3.8$ & $25.5\pm4.2$\\
8784.65645 & $23.2\pm3.6$ & $37.9\pm4.1$\\
8787.65939 & $23.2\pm3.1$ & $31.7\pm3.6$\\
8788.67136 & $26.3\pm3.8$ & $34.8\pm4.2$\\
8797.67104 & $30.9\pm3.1$ & $37.1\pm3.6$\\
\hline
\end{tabular}
\end{table}
\end{center}

\section{Discussion}

According to the orbital solution, derived with our STK astrometry, comet 2I/Borisov passes through its perihelion ($q=2.0066$\,au) on 2019 Dec 8.6 and reaches its closest distance to Earth ($\Delta_{\text{min}}=1.9368$\,au) on 2019 Dec 28.2. The orbit of the comet is found to be significantly hyperbolic ($e=3.3570 \pm 0.0006$) and inclined to the ecliptic ($i=44.0524 \pm 0.0004\,^{\circ}$). Hence, we can confirm the interstellar origin of the comet, because very close encounters with any of the giant planets in the solar system, which might have induced the high orbital eccentricity of the comet, can clearly be ruled out before its perihelion passage.

The residuals of the derived orbital solution of comet 2I/Borisov are listed in Tab.\,\ref{astrometry}, and their histograms are shown in Fig.\,\ref{residuals}. The residuals are normal distributed (confirmed e.g. with Anderson-Darling tests) as expected, because the RMS of the orbital solution well agrees with the average accuracy of the STK astrometry of the comet. In particular, there are no systematic deviations detected between the derived orbital solution and the STK astrometry of the comet.

In addition, we have also tested the astrometric motion of comet 2I/Borisov for non-gravitational effects, as expected in comets due to the evaporation of their icy surfaces when approaching the Sun. For this purpose, we have included in the orbit fitting non-gravitational accelerations in the anti-solar and transversal directions, described by the parameters $A_1$ and $A_2$. The orbital elements of comet 2I/Borisov, including non-gravitational effects, were derived again with \href{https://www.projectpluto.com/find_orb.htm}{\textsc{find\_orb}}, and are summarized in Tab.\,\ref{nongrav_elements}\hspace{-1.5mm}. The orbital solution with included non-gravitational accelerations exhibits a $RMS=0.348$\,arcsec, which is only slightly smaller than the RMS of the standard orbital solution, taking into account only gravitational perturbers (see above). The orbital elements of both solutions do not significantly differ from each other, and the obtained non-gravitational accelerations of the comet are $A_1 = (-7.2 \pm 3.96) \cdot 10^{-7}$\,au/day$^{2}$, and $A_2 = (1.04 \pm 0.45) \cdot 10^{-6}$\,au/day$^{2}$, respectively. Therefore, we conclude that there is no statistical evidence ($SNR > 3$) for non-gravitational effects that have significantly altered the motion of comet 2I/Borisov during our observing campaign.

\begin{center}
\begin{table}[h]
\caption{Orbital elements of comet 2I/Borisov (for heliocentric ecliptic J\,2000) of the best-fitting orbital solution, including non-gravitational effects.\label{nongrav_elements}}
\centering
\begin{tabular}{ll}
\hline
Epoch    & 2019 Nov 1 (JD 2458788.5)\\
\hline
$T$               & $2019 \text{~Dec~} 8.52 \pm 0.02$\\
$q [\text{au}]$   & $2.0075 \pm 0.0004$\\
$e$               & $3.362 \pm 0.002$\\
$i~[^\circ]$      & $44.044 \pm 0.004$\\
$\omega~[^\circ]$ & $209.10 \pm 0.02$\\
$\Omega~[^\circ]$ & $308.155 \pm 0.003$\\
\hline
$A_{1}~[\text{au}/\text{day}^2]$  & $(-7.20\pm3.96)\cdot10^{-7}$\\
$A_{2}~[\text{au}/\text{day}^2]$  & $(1.04\pm0.45)\cdot10^{-6}$\\
\hline
\end{tabular}
\end{table}
\end{center}

\begin{figure}[h]
\centerline{\includegraphics[width=0.49\textwidth]{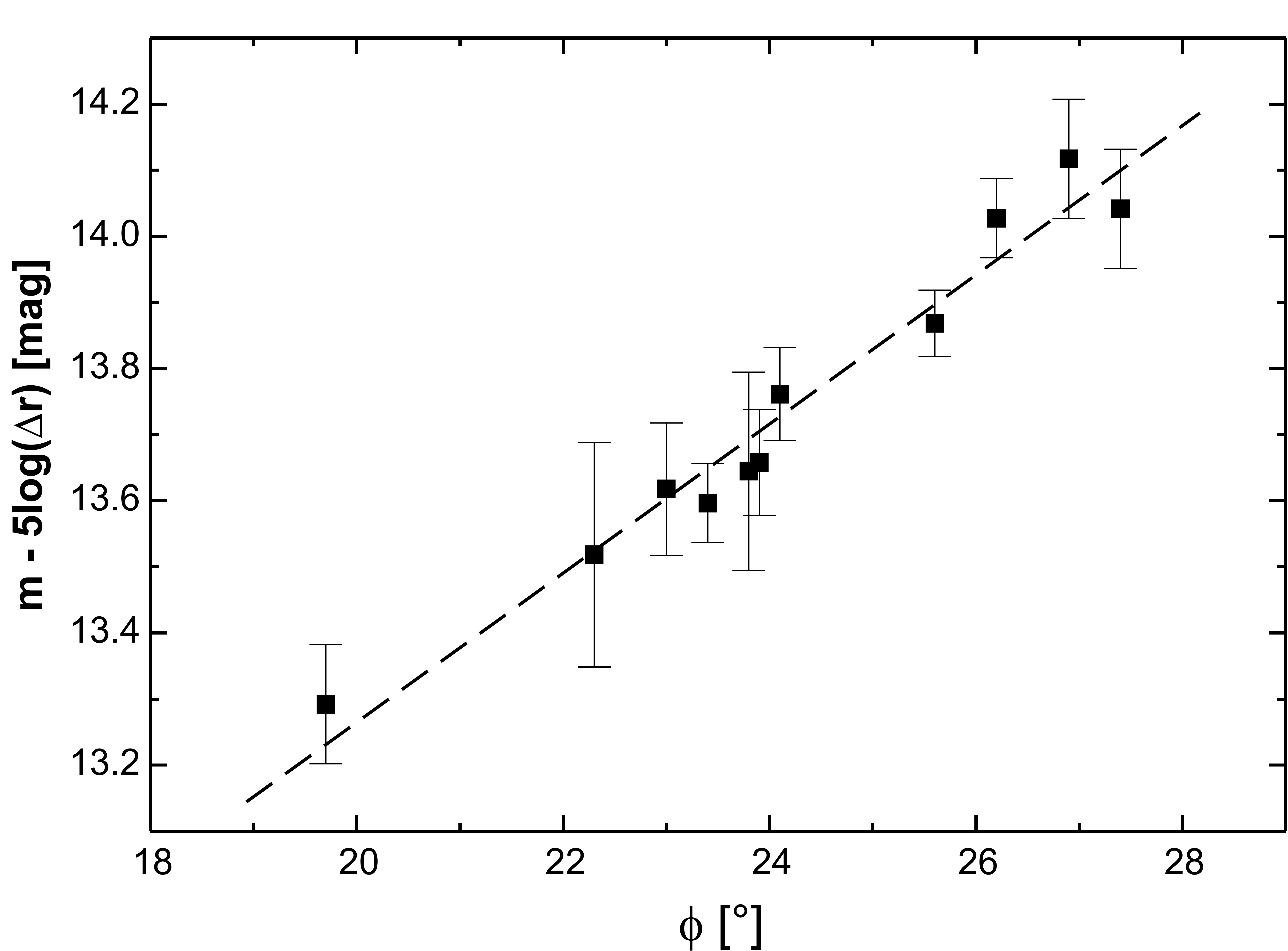}}
\caption{The distance corrected G-band magnitude of the nucleus of comet 2I/Borisov for all STK observing epochs, plotted versus its phase angle $\phi$, as determined with the best-fitting orbital solution of the comet, based on our STK astrometry. The distance-corrected photometry of the nucleus clearly exhibits a linear correlation with its phase angle (dashed line).}\label{photocorr}
\end{figure}

The STK photometry of the nucleus of comet 2I/Borisov, as measured in our STK images, is found to be stable throughout our observing campaign, although the location of the comet relative to the Sun and Earth has significantly changed between our first and last observing epoch. The distance-corrected photometry of the nucleus of comet 2I/Borisov is shown in Fig.\,\ref{photocorr} for all observing epochs, dependant on its phase angle $\phi$. The distances $r$ to Sun and $\Delta$ to Earth, as well as the phase angle $\phi$, are all obtained from our best-fitting orbital solution of the comet. The distance-corrected photometry of the nucleus of comet 2I/Borisov clearly exhibits a linear correlation with its phase angle (correlation coefficient of $R=0.975$, and $P<0.0001$ for the t-test of zero-slope), which is indicated with a dashed line in Fig.\,\ref{photocorr}\hspace{-1.5mm}.

According to \cite{jewitt1988}, the apparent magnitude $m$ of the nucleus of a comet, which is observed from the distance $\Delta$ while being located at a solar distance $r$, is described by

\begin{equation}
m = H + 5\log(\Delta r) + \Phi
\end{equation}\newline
with its absolute magnitude $H$, i.e. the apparent magnitude of the nucleus, as seen from the Sun at a distance of 1\,au, and the phase function $\Phi=\beta\phi$, with $\beta$ the linear phase function coefficient.

The linear fitting of the distance-corrected STK photometry of the nucleus of comet 2I/Borisov as a function of its phase angle, yields an absolute magnitude of $H=11.1\pm0.2$\,mag, and a linear phase function coefficient $\beta=0.111\pm0.008$\,mag/deg.

The radius of the nucleus can be derived with its absolute magnitude $H$, its geometric albedo $A$, as well as the apparent magnitude of the Sun (here in the G-band) $m_\odot=-26.9$\,mag \citep{casagrande2018}

\begin{equation}
R = 1\,\text{au}~\frac{10^{( m_\odot - H ) / 5}}{\sqrt{A}}
\end{equation}\newline
Assuming a geometric albedo of $A=0.04$, which corresponds to the median value found for comets of the Jupiter family \citep{kokotanekova2017}, we obtain for the nucleus of comet 2I/Borisov a radius of $R=18.8\pm1.7$\,km. This result should, however, be considered as an upper limit of the radius because the geometric albedo of the nucleus could be larger, and furthermore the comet was active during our imaging campaign. On the one hand, the activity of the nucleus of the comet is indicated by its phase function coefficient, which is steeper than the one expected for an inactive core of a comet, e.g. $\beta=0.046$\,mag/deg, which is the median value of inactive comets of the Jupiter-family \citep{kokotanekova2017}. On the other hand, the activity of the nucleus can also be seen in the morphology of the comet, which was recorded in our deep STK images. The coma and the tail of comet 2I/Borisov are both well resolved in these images, which exhibit extents (detected at $SNR \ge 5$) that significantly exceed the $FWHM$ of detected background stars (i.e. of point spread functions).

Finally, the analysis of the morphology of comet 2I/Borisov allows the characterization of the geometry of its coma and tail. The diameter $\text{D}_{\text{Coma}}$ of the coma of the comet can be determined from its apparent diameter $\varnothing_{\text{Coma}}$, measured in our deep STK images, and its distance $\Delta$ to Earth during the observations, known from the derived best-fitting orbital solution of the comet

\begin{equation}
\text{D}_{\text{Coma}}=2 \Delta \sin \left( \frac{1}{2} \varnothing_{\text{Coma}} \right)
\end{equation}\newline
By adopting an anti-solar orientation, the length of the tail of the comet can be derived with its apparent length $\Lambda_{\text{Tail}}$, measured in our deep STK images, the distance $\Delta$ of the comet to Earth, as well as its phase angle $\phi$ during the observations, both determined with the derived best-fitting orbital solution of the comet

\begin{equation}
\text{L}_{\text{Tail}} = \frac{ \Delta \sin( \Lambda_{\text{Tail}} )}{ \sin( \phi - \Lambda_{\text{Tail}} )}
\end{equation}\newline
The obtained diameter of the coma of comet 2I/Borisov, and the length of its tail, are summarized for all deep STK imaging observations in Tab.\,\ref{morph}\hspace{-1.5mm}. The diameter of the coma, as well as the length of the tail of the comet, do not significantly vary within the span of time, covered by our deep STK imaging observations. On average, the comet exhibits a coma with a diameter of $(4.57\pm0.38)\cdot10^4$\,km and a tail with a length of $(1.52\pm0.12)\cdot10^5$\,km, assuming that it is orientated in the anti-solar direction.

\begin{center}
\begin{table}[h]
\caption{The diameter of the coma of comet 2I/Borisov and the length of its tail, derived from all deep STK imaging observations, taken during our observing campaign of the comet.}\label{morph}
\centering
\begin{tabular}{ccc}
\hline
JD$-$2450000 & $\text{D}_{\text{Coma}}~[10^4~\text{km}]$ & $\text{L}_{\text{Tail}}~[10^5~\text{km}]$ \\
\hline
8765.63414   & $4.80 \pm 0.79$           & $1.57 \pm 0.26$\\
8784.65645   & $4.20 \pm 0.65$           & $1.73 \pm 0.19$\\
8787.65939   & $4.11 \pm 0.55$           & $1.39 \pm 0.16$\\
8788.67136   & $4.63 \pm 0.67$           & $1.50 \pm 0.18$\\
8797.67104   & $5.13 \pm 0.51$           & $1.43 \pm 0.14$\\
\hline
\end{tabular}
\end{table}
\end{center}

Further observations of comet 2I/Borisov, which have to be carried out preferably from the southern hemisphere, will show how the activity of the comet will evolve, especially after its perihelion passage. According to our orbital solution and photometric analysis of comet 2I/Borisov, after its perihelion passage on 2019 Dec 8 the comet will always be located on the sky southward of $\text{Dec}=-18\,^\circ$, and the apparent brightness of its nucleus should continuously decrease after January 2020. However, follow-up observations of the comet with a few minutes of integration time should be feasible until mid of May 2020 at 1\,m class telescopes, which are equipped with state-of-the-art CCD detectors.

\section*{Acknowledgments}

We would like to thank A. Gonzalez, T. Heyne, F. Hildebrandt, E. Hohmann, A. Nowotnick, J. Tietz, and E. Wahl for their assistance during some observations. We made use of data taken by the European Space Agency (ESA) mission {\it Gaia} (https://www.cosmos.esa.int/gaia), processed by the {\it Gaia} Data Processing and Analysis Consortium (DPAC, https://www.cosmos.esa.int/web/gaia/dpac/consortium). Funding for the DPAC has been provided by national institutions, in particular the institutions participating in the {\it Gaia} Multilateral Agreement.

\bibliography{mugrauer}

\newpage

\appendix

\section{STK deep imaging Observations and Astrometry of comet 2I/Borisov}

\begin{figure*}[h]
\centerline{\includegraphics[width=0.737\textwidth]{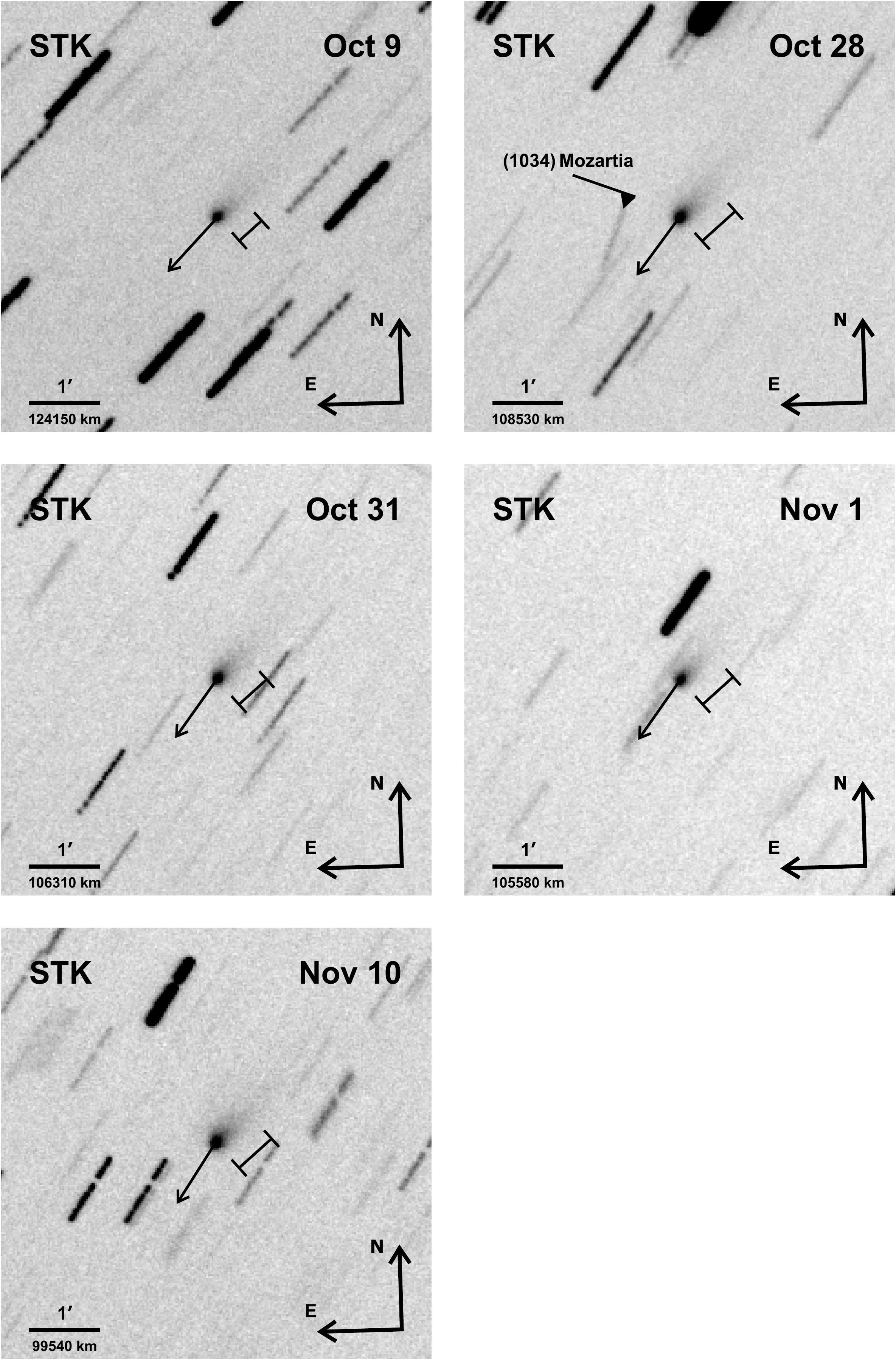}}
\caption{Deep STK images of comet 2I/Borisov, taken in dark time and during clear sky conditions in 5 observing epochs in October and November 2019. These images are the averages of STK frames, taken in each observing epoch, which were shifted before the averaging to correct for the motion of the comet during the observations. The direction of the motion of the comet is indicated with a black arrow. On Oct 28 the asteroid (1034) Mozartia, named after Wolfgang Amadeus Mozart, is detected about 1\,arcmin east of comet 2I/Borisov. Due to its proper motion, the asteroid appears as a trail, whose orientation differs from that of the background stars. The image scaling is shown in each deep STK image, determined with the pixel scale of the STK detector and the distance $\Delta$ of the comet to Earth during the observations, derived with the best-fitting orbital solution of the comet. Both coma and tail of comet 2I/Borisov are clearly detected in all deep STK images. The apparent length of the tail of the comet (detected at $SNR \ge 5$) is illustrated with length indicators in all images.}\label{pics}
\end{figure*}

\begin{center}
\begin{table*}[h]
\caption{STK astrometry of comet 2I/Borisov. The equatorial coordinates of the comet, measured in the STK images, are listed together with the residuals of the best-fitting orbital solution of the comet, derived with these astrometric measurements.}\label{astrometry}
\centering
\begin{tabular}{ccccc}
\hline
Date & RA & Dec & $\text{RA}\,(\text{O}-\text{C})$ & $\text{Dec}\,(\text{O}-\text{C})$\\
yyyy mm dd.ddddd & $[$hh mm ss.sss$]$ & $[$dd mm ss.ss$]$ & $[$arcsec$]$ & $[$arcsec$]$\\
\hline
2019 10 09.11796 & 09 41 24.480 & $+$19 42 18.04 & $+$0.08 & $-$0.26\\
2019 10 09.11950 & 09 41 24.645 & $+$19 42 15.58 & $+$0.02 & $-$0.24\\
2019 10 09.12103 & 09 41 24.842 & $+$19 42 12.91 & $+$0.42 & $-$0.45\\
2019 10 09.12257 & 09 41 24.977 & $+$19 42 10.89 & $-$0.08 & $+$0.01\\
2019 10 09.12411 & 09 41 25.148 & $+$19 42 08.27 & $-$0.06 & $-$0.14\\
2019 10 09.12564 & 09 41 25.305 & $+$19 42 05.75 & $-$0.22 & $-$0.19\\
2019 10 09.12718 & 09 41 25.495 & $+$19 42 03.53 & $+$0.06 & $+$0.06\\
2019 10 09.12870 & 09 41 25.651 & $+$19 42 00.73 & $-$0.10 & $-$0.29\\
2019 10 09.13024 & 09 41 25.831 & $+$19 41 58.09 & $+$0.04 & $-$0.45\\
2019 10 09.13177 & 09 41 25.985 & $+$19 41 55.83 & $-$0.16 & $-$0.25\\
2019 10 09.13331 & 09 41 26.184 & $+$19 41 53.20 & $+$0.25 & $-$0.40\\
2019 10 09.13484 & 09 41 26.306 & $+$19 41 51.32 & $-$0.41 & $+$0.18\\
2019 10 09.13638 & 09 41 26.494 & $+$19 41 49.23 & $-$0.15 & $+$0.57\\
2019 10 09.13791 & 09 41 26.661 & $+$19 41 46.04 & $-$0.17 & $-$0.16\\
2019 10 09.13944 & 09 41 26.861 & $+$19 41 43.54 & $+$0.27 & $-$0.20\\
2019 10 09.14097 & 09 41 26.979 & $+$19 41 41.43 & $-$0.44 & $+$0.15\\
2019 10 09.14251 & 09 41 27.189 & $+$19 41 38.45 & $+$0.13 & $-$0.35\\
2019 10 09.14419 & 09 41 27.380 & $+$19 41 36.49 & $+$0.21 & $+$0.39\\
2019 10 09.14573 & 09 41 27.514 & $+$19 41 33.46 & $-$0.29 & $-$0.16\\
2019 10 09.14726 & 09 41 27.723 & $+$19 41 30.55 & $+$0.28 & $-$0.60\\
2019 10 09.14880 & 09 41 27.854 & $+$19 41 28.59 & $-$0.27 & $-$0.09\\
2019 10 09.15032 & 09 41 28.023 & $+$19 41 26.76 & $-$0.25 & $+$0.53\\
2019 10 22.13286 & 10 05 13.746 & $+$13 22 03.67 & $+$0.03 & $-$0.67\\
2019 10 22.13372 & 10 05 13.799 & $+$13 22 01.95 & $-$0.56 & $-$0.76\\
2019 10 22.13456 & 10 05 13.870 & $+$13 22 01.27 & $-$0.86 & $+$0.15\\
2019 10 22.13542 & 10 05 14.066 & $+$13 22 00.14 & $+$0.64 & $+$0.65\\
2019 10 22.13626 & 10 05 14.104 & $+$13 21 58.52 & $-$0.14 & $+$0.62\\
2019 10 22.13878 & 10 05 14.424 & $+$13 21 53.25 & $+$0.54 & $+$0.13\\
2019 10 22.15066 & 10 05 15.656 & $+$13 21 30.66 & $-$0.31 & $+$0.06\\
2019 10 22.15150 & 10 05 15.773 & $+$13 21 29.45 & $+$0.07 & $+$0.44\\
2019 10 22.15234 & 10 05 15.881 & $+$13 21 28.20 & $+$0.32 & $+$0.79\\
2019 10 22.15318 & 10 05 15.971 & $+$13 21 26.20 & $+$0.30 & $+$0.38\\
2019 10 22.15402 & 10 05 16.029 & $+$13 21 24.44 & $-$0.19 & $+$0.21\\
2019 10 22.15486 & 10 05 16.098 & $+$13 21 23.37 & $-$0.51 & $+$0.73\\
2019 10 22.15569 & 10 05 16.240 & $+$13 21 20.82 & $+$0.25 & $-$0.24\\
2019 10 22.15654 & 10 05 16.291 & $+$13 21 19.29 & $-$0.35 & $-$0.16\\
2019 10 22.15737 & 10 05 16.399 & $+$13 21 18.09 & $-$0.09 & $+$0.21\\
2019 10 22.15822 & 10 05 16.486 & $+$13 21 15.81 & $-$0.17 & $-$0.45\\
2019 10 22.15906 & 10 05 16.569 & $+$13 21 14.74 & $-$0.29 & $+$0.07\\
2019 10 22.15991 & 10 05 16.699 & $+$13 21 13.69 & $+$0.27 & $+$0.63\\
2019 10 22.16076 & 10 05 16.788 & $+$13 21 11.15 & $+$0.22 & $-$0.30\\
2019 10 22.16161 & 10 05 16.875 & $+$13 21 10.15 & $+$0.14 & $+$0.31\\
2019 10 22.16245 & 10 05 16.988 & $+$13 21 07.90 & $+$0.46 & $-$0.34\\
2019 10 22.16330 & 10 05 17.006 & $+$13 21 06.96 & $-$0.62 & $+$0.33\\
2019 10 22.16414 & 10 05 17.144 & $+$13 21 04.57 & $+$0.06 & $-$0.47\\
\hline
\end{tabular}
\end{table*}
\end{center}

\setcounter{table}{0}
\begin{center}
\begin{table*}[h]
\caption{continued}
\centering
\begin{tabular}{ccccc}
\hline
Date & RA & Dec & $\text{RA}\,(\text{O}-\text{C})$ & $\text{Dec}\,(\text{O}-\text{C})$\\
yyyy mm dd.ddddd & $[$hh mm ss.sss$]$ & $[$dd mm ss.ss$]$ & $[$arcsec$]$ & $[$arcsec$]$\\
\hline
2019 10 22.16499 & 10 05 17.271 & $+$13 21 03.65 & $+$0.57 & $+$0.22\\
2019 10 22.16583 & 10 05 17.319 & $+$13 21 01.93 & $-$0.06 & $+$0.10\\
2019 10 26.14944 & 10 12 30.760 & $+$11 11 53.69 & $+$0.30 & $-$0.57\\
2019 10 26.15097 & 10 12 30.876 & $+$11 11 52.09 & $-$0.43 & $+$0.87\\
2019 10 26.15405 & 10 12 31.269 & $+$11 11 45.36 & $+$0.46 & $+$0.26\\
2019 10 26.15865 & 10 12 31.694 & $+$11 11 35.56 & $-$0.59 & $-$0.39\\
2019 10 28.14111 & 10 16 06.608 & $+$10 05 02.23 & $+$0.05 & $-$0.20\\
2019 10 28.14265 & 10 16 06.794 & $+$10 04 58.96 & $+$0.35 & $-$0.34\\
2019 10 28.14418 & 10 16 06.944 & $+$10 04 55.76 & $+$0.13 & $-$0.42\\
2019 10 28.14572 & 10 16 07.087 & $+$10 04 53.39 & $-$0.21 & $+$0.34\\
2019 10 28.14725 & 10 16 07.270 & $+$10 04 49.75 & $+$0.06 & $-$0.19\\
2019 10 28.14877 & 10 16 07.453 & $+$10 04 46.45 & $+$0.35 & $-$0.39\\
2019 10 28.15031 & 10 16 07.586 & $+$10 04 43.58 & $-$0.14 & $-$0.13\\
2019 10 28.15184 & 10 16 07.782 & $+$10 04 40.66 & $+$0.32 & $+$0.07\\
2019 10 28.15338 & 10 16 07.909 & $+$10 04 37.81 & $-$0.25 & $+$0.35\\
2019 10 28.15491 & 10 16 08.111 & $+$10 04 34.78 & $+$0.30 & $+$0.44\\
2019 10 28.15645 & 10 16 08.238 & $+$10 04 31.35 & $-$0.27 & $+$0.15\\
2019 10 28.15799 & 10 16 08.401 & $+$10 04 27.84 & $-$0.31 & $-$0.23\\
2019 10 28.15951 & 10 16 08.607 & $+$10 04 25.35 & $+$0.31 & $+$0.38\\
2019 10 28.16105 & 10 16 08.765 & $+$10 04 21.46 & $+$0.20 & $-$0.38\\
2019 10 28.16258 & 10 16 08.946 & $+$10 04 18.25 & $+$0.44 & $-$0.47\\
2019 10 28.16412 & 10 16 09.036 & $+$10 04 16.47 & $-$0.68 & $+$0.88\\
2019 10 28.16566 & 10 16 09.239 & $+$10 04 12.87 & $-$0.12 & $+$0.42\\
2019 10 28.16719 & 10 16 09.408 & $+$10 04 10.12 & $-$0.06 & $+$0.78\\
2019 10 28.16873 & 10 16 09.572 & $+$10 04 06.60 & $-$0.08 & $+$0.40\\
2019 10 28.17027 & 10 16 09.778 & $+$10 04 03.20 & $+$0.51 & $+$0.14\\
2019 10 28.17179 & 10 16 09.891 & $+$10 04 00.22 & $-$0.23 & $+$0.25\\
2019 10 30.16708 & 10 19 45.549 & $+$08 55 26.54 & $-$0.20 & $+$0.15\\
2019 10 30.16862 & 10 19 45.696 & $+$08 55 22.89 & $-$0.47 & $-$0.29\\
2019 10 30.17016 & 10 19 45.872 & $+$08 55 19.84 & $-$0.31 & $-$0.13\\
2019 10 30.17169 & 10 19 46.097 & $+$08 55 16.09 & $+$0.59 & $-$0.69\\
2019 10 30.17323 & 10 19 46.260 & $+$08 55 13.50 & $+$0.56 & $-$0.07\\
2019 10 30.19561 & 10 19 48.632 & $+$08 54 27.22 & $+$0.16 & $+$0.29\\
2019 10 30.19715 & 10 19 48.772 & $+$08 54 24.11 & $-$0.21 & $+$0.39\\
2019 10 30.19869 & 10 19 48.985 & $+$08 54 20.37 & $+$0.50 & $-$0.14\\
2019 10 30.20022 & 10 19 49.121 & $+$08 54 17.11 & $+$0.09 & $-$0.21\\
2019 10 30.20176 & 10 19 49.253 & $+$08 54 14.63 & $-$0.40 & $+$0.52\\
2019 10 31.14708 & 10 21 31.309 & $+$08 21 11.45 & $+$0.54 & $-$0.35\\
2019 10 31.14862 & 10 21 31.482 & $+$08 21 08.57 & $+$0.66 & $+$0.02\\
2019 10 31.15015 & 10 21 31.632 & $+$08 21 05.41 & $+$0.45 & $+$0.08\\
2019 10 31.15168 & 10 21 31.741 & $+$08 21 02.75 & $-$0.36 & $+$0.64\\
2019 10 31.15476 & 10 21 32.108 & $+$08 20 55.69 & $+$0.18 & $+$0.07\\
2019 10 31.15630 & 10 21 32.258 & $+$08 20 52.31 & $-$0.04 & $-$0.07\\
2019 10 31.15785 & 10 21 32.414 & $+$08 20 49.06 & $-$0.19 & $-$0.05\\
2019 10 31.15939 & 10 21 32.598 & $+$08 20 45.79 & $+$0.09 & $-$0.08\\
\hline
\end{tabular}
\end{table*}
\end{center}

\setcounter{table}{0}
\begin{center}
\begin{table*}[h]
\caption{continued}
\centering
\begin{tabular}{ccccc}
\hline
Date & RA & Dec & $\text{RA}\,(\text{O}-\text{C})$ & $\text{Dec}\,(\text{O}-\text{C})$\\
yyyy mm dd.ddddd & $[$hh mm ss.sss$]$ & $[$dd mm ss.ss$]$ & $[$arcsec$]$ & $[$arcsec$]$\\
\hline
2019 10 31.16093 & 10 21 32.751 & $+$08 20 42.80 & $-$0.08 & $+$0.18\\
2019 10 31.16247 & 10 21 32.928 & $+$08 20 39.44 & $+$0.09 & $+$0.06\\
2019 10 31.16400 & 10 21 33.084 & $+$08 20 36.28 & $-$0.02 & $+$0.13\\
2019 10 31.16554 & 10 21 33.262 & $+$08 20 32.69 & $+$0.17 & $-$0.22\\
2019 10 31.16708 & 10 21 33.412 & $+$08 20 30.14 & $-$0.05 & $+$0.48\\
2019 10 31.16862 & 10 21 33.576 & $+$08 20 26.60 & $-$0.06 & $+$0.18\\
2019 10 31.17015 & 10 21 33.749 & $+$08 20 23.03 & $+$0.07 & $-$0.16\\
2019 10 31.17169 & 10 21 33.902 & $+$08 20 20.20 & $-$0.10 & $+$0.25\\
2019 11 01.14683 & 10 23 18.987 & $+$07 45 52.44 & $+$0.59 & $-$0.21\\
2019 11 01.14837 & 10 23 19.122 & $+$07 45 49.08 & $+$0.14 & $-$0.29\\
2019 11 01.14991 & 10 23 19.263 & $+$07 45 45.70 & $-$0.21 & $-$0.39\\
2019 11 01.15145 & 10 23 19.462 & $+$07 45 42.90 & $+$0.30 & $+$0.09\\
2019 11 01.15299 & 10 23 19.582 & $+$07 45 39.43 & $-$0.37 & $-$0.10\\
2019 11 01.15453 & 10 23 19.777 & $+$07 45 36.76 & $+$0.08 & $+$0.51\\
2019 11 01.15606 & 10 23 19.976 & $+$07 45 33.09 & $+$0.61 & $+$0.10\\
2019 11 01.15760 & 10 23 20.111 & $+$07 45 29.66 & $+$0.17 & $-$0.05\\
2019 11 01.15914 & 10 23 20.237 & $+$07 45 26.31 & $-$0.41 & $-$0.12\\
2019 11 01.16068 & 10 23 20.436 & $+$07 45 23.76 & $+$0.10 & $+$0.62\\
2019 11 01.16222 & 10 23 20.599 & $+$07 45 20.56 & $+$0.07 & $+$0.70\\
2019 11 01.16376 & 10 23 20.751 & $+$07 45 17.06 & $-$0.11 & $+$0.48\\
2019 11 01.16530 & 10 23 20.960 & $+$07 45 13.06 & $+$0.54 & $-$0.24\\
2019 11 01.16683 & 10 23 21.122 & $+$07 45 09.64 & $+$0.52 & $-$0.40\\
2019 11 01.16837 & 10 23 21.270 & $+$07 45 06.00 & $+$0.27 & $-$0.76\\
2019 11 01.16990 & 10 23 21.438 & $+$07 45 03.25 & $+$0.34 & $-$0.25\\
2019 11 01.17756 & 10 23 22.233 & $+$07 44 46.79 & $-$0.02 & $-$0.39\\
2019 11 01.17910 & 10 23 22.426 & $+$07 44 43.59 & $+$0.41 & $-$0.31\\
2019 11 01.18063 & 10 23 22.569 & $+$07 44 39.94 & $+$0.10 & $-$0.70\\
2019 11 01.18215 & 10 23 22.743 & $+$07 44 37.48 & $+$0.27 & $+$0.08\\
2019 11 01.18369 & 10 23 22.917 & $+$07 44 33.62 & $+$0.41 & $-$0.50\\
2019 11 01.18524 & 10 23 23.021 & $+$07 44 30.57 & $-$0.50 & $-$0.24\\
2019 11 01.18677 & 10 23 23.223 & $+$07 44 27.29 & $+$0.07 & $-$0.26\\
2019 11 01.18831 & 10 23 23.362 & $+$07 44 24.79 & $-$0.31 & $+$0.52\\
2019 11 01.18984 & 10 23 23.583 & $+$07 44 20.95 & $+$0.54 & $-$0.06\\
2019 11 01.19138 & 10 23 23.680 & $+$07 44 17.93 & $-$0.46 & $+$0.20\\
2019 11 01.19291 & 10 23 23.919 & $+$07 44 14.24 & $+$0.66 & $-$0.23\\
2019 11 01.19444 & 10 23 24.030 & $+$07 44 11.62 & $-$0.12 & $+$0.41\\
2019 11 01.19598 & 10 23 24.246 & $+$07 44 08.01 & $+$0.65 & $+$0.09\\
2019 11 10.15568 & 10 39 22.114 & $+$02 10 10.23 & $-$0.36 & $+$0.27\\
2019 11 10.15722 & 10 39 22.304 & $+$02 10 06.56 & $+$0.05 & $+$0.20\\
2019 11 10.15875 & 10 39 22.433 & $+$02 10 02.61 & $-$0.43 & $-$0.17\\
2019 11 10.16029 & 10 39 22.627 & $+$02 09 59.20 & $+$0.04 & $+$0.02\\
2019 11 10.16182 & 10 39 22.760 & $+$02 09 55.96 & $-$0.38 & $+$0.36\\
2019 11 10.16334 & 10 39 22.947 & $+$02 09 51.77 & $+$0.02 & $-$0.28\\
2019 11 10.16488 & 10 39 23.095 & $+$02 09 48.78 & $-$0.20 & $+$0.33\\
2019 11 10.16641 & 10 39 23.284 & $+$02 09 44.39 & $+$0.22 & $-$0.48\\
\hline
\end{tabular}
\end{table*}
\end{center}

\setcounter{table}{0}
\begin{center}
\begin{table*}[h]
\caption{continued}
\centering
\begin{tabular}{ccccc}
\hline
Date & RA & Dec & $\text{RA}\,(\text{O}-\text{C})$ & $\text{Dec}\,(\text{O}-\text{C})$\\
yyyy mm dd.ddddd & $[$hh mm ss.sss$]$ & $[$dd mm ss.ss$]$ & $[$arcsec$]$ & $[$arcsec$]$\\
\hline
2019 11 10.16795 & 10 39 23.426 & $+$02 09 41.22 & $-$0.09 & $-$0.05\\
2019 11 10.16948 & 10 39 23.587 & $+$02 09 37.83 & $-$0.09 & $+$0.14\\
2019 11 10.17101 & 10 39 23.759 & $+$02 09 33.66 & $+$0.07 & $-$0.45\\
2019 11 10.17255 & 10 39 23.935 & $+$02 09 30.58 & $+$0.28 & $+$0.07\\
2019 11 10.17407 & 10 39 24.079 & $+$02 09 26.38 & $+$0.03 & $-$0.58\\
2019 11 10.17561 & 10 39 24.259 & $+$02 09 23.76 & $+$0.30 & $+$0.41\\
2019 11 10.17714 & 10 39 24.410 & $+$02 09 19.62 & $+$0.15 & $-$0.15\\
2019 11 10.18485 & 10 39 25.218 & $+$02 09 02.06 & $+$0.08 & $+$0.32\\
2019 11 10.18639 & 10 39 25.368 & $+$02 08 58.46 & $-$0.10 & $+$0.32\\
2019 11 14.16098 & 10 46 26.158 & $-$00 28 54.11 & $-$0.67 & $-$0.10\\
2019 11 14.16251 & 10 46 26.337 & $-$00 28 57.89 & $-$0.38 & $-$0.17\\
2019 11 14.16405 & 10 46 26.521 & $-$00 29 01.75 & $-$0.04 & $-$0.29\\
2019 11 14.16711 & 10 46 26.826 & $-$00 29 08.80 & $-$0.27 & $+$0.08\\
2019 11 14.16865 & 10 46 26.992 & $-$00 29 12.81 & $-$0.20 & $-$0.19\\
2019 11 19.16215 & 10 55 11.966 & $-$03 55 25.64 & $+$0.78 & $-$0.22\\
2019 11 19.16368 & 10 55 12.130 & $-$03 55 29.46 & $+$0.85 & $-$0.17\\
2019 11 19.16522 & 10 55 12.247 & $-$03 55 33.37 & $+$0.21 & $-$0.19\\
2019 11 19.16675 & 10 55 12.365 & $-$03 55 36.87 & $-$0.40 & $+$0.18\\
2019 11 19.16981 & 10 55 12.738 & $-$03 55 45.22 & $+$0.43 & $-$0.43\\
2019 11 19.17134 & 10 55 12.890 & $-$03 55 49.06 & $+$0.33 & $-$0.40\\
2019 11 19.17288 & 10 55 13.021 & $-$03 55 52.48 & $-$0.10 & $+$0.07\\
2019 11 19.17448 & 10 55 13.161 & $-$03 55 56.45 & $-$0.49 & $+$0.15\\
2019 11 19.17602 & 10 55 13.358 & $-$03 56 00.66 & $+$0.07 & $-$0.17\\
2019 11 19.17755 & 10 55 13.518 & $-$03 56 04.95 & $+$0.09 & $-$0.59\\
2019 11 23.18795 & 11 02 11.687 & $-$06 47 30.32 & $-$0.38 & $+$0.16\\
2019 11 23.18949 & 11 02 11.886 & $-$06 47 34.24 & $+$0.22 & $+$0.24\\
2019 11 23.19102 & 11 02 12.038 & $-$06 47 38.90 & $+$0.14 & $-$0.44\\
2019 11 23.19256 & 11 02 12.206 & $-$06 47 42.15 & $+$0.28 & $+$0.32\\
2019 11 23.19410 & 11 02 12.335 & $-$06 47 46.96 & $-$0.15 & $-$0.48\\
2019 11 23.19564 & 11 02 12.454 & $-$06 47 49.98 & $-$0.74 & $+$0.50\\
2019 11 23.19718 & 11 02 12.614 & $-$06 47 54.22 & $-$0.72 & $+$0.27\\
2019 11 23.19872 & 11 02 12.839 & $-$06 47 58.02 & $+$0.28 & $+$0.47\\
2019 11 23.20025 & 11 02 12.990 & $-$06 48 02.33 & $+$0.18 & $+$0.14\\
2019 11 23.20179 & 11 02 13.127 & $-$06 48 06.14 & $-$0.14 & $+$0.34\\
2019 11 23.20345 & 11 02 13.337 & $-$06 48 11.13 & $+$0.45 & $-$0.33\\
2019 11 23.20499 & 11 02 13.474 & $-$06 48 14.53 & $+$0.13 & $+$0.27\\
2019 11 23.20653 & 11 02 13.625 & $-$06 48 18.64 & $+$0.02 & $+$0.17\\
2019 11 23.20807 & 11 02 13.813 & $-$06 48 23.47 & $+$0.46 & $-$0.65\\
2019 11 23.20961 & 11 02 13.918 & $-$06 48 26.74 & $-$0.33 & $+$0.08\\
2019 11 23.21115 & 11 02 14.121 & $-$06 48 30.70 & $+$0.33 & $+$0.13\\
2019 11 23.21269 & 11 02 14.233 & $-$06 48 34.29 & $-$0.36 & $+$0.54\\
2019 11 23.21422 & 11 02 14.381 & $-$06 48 38.30 & $-$0.50 & $+$0.51\\
2019 11 23.21576 & 11 02 14.590 & $-$06 48 42.40 & $+$0.26 & $+$0.42\\
2019 11 23.21730 & 11 02 14.721 & $-$06 48 46.63 & $-$0.15 & $+$0.20\\
\hline
\end{tabular}
\end{table*}
\end{center}

\begin{figure*}[h]
\centerline{\includegraphics[width=1\textwidth]{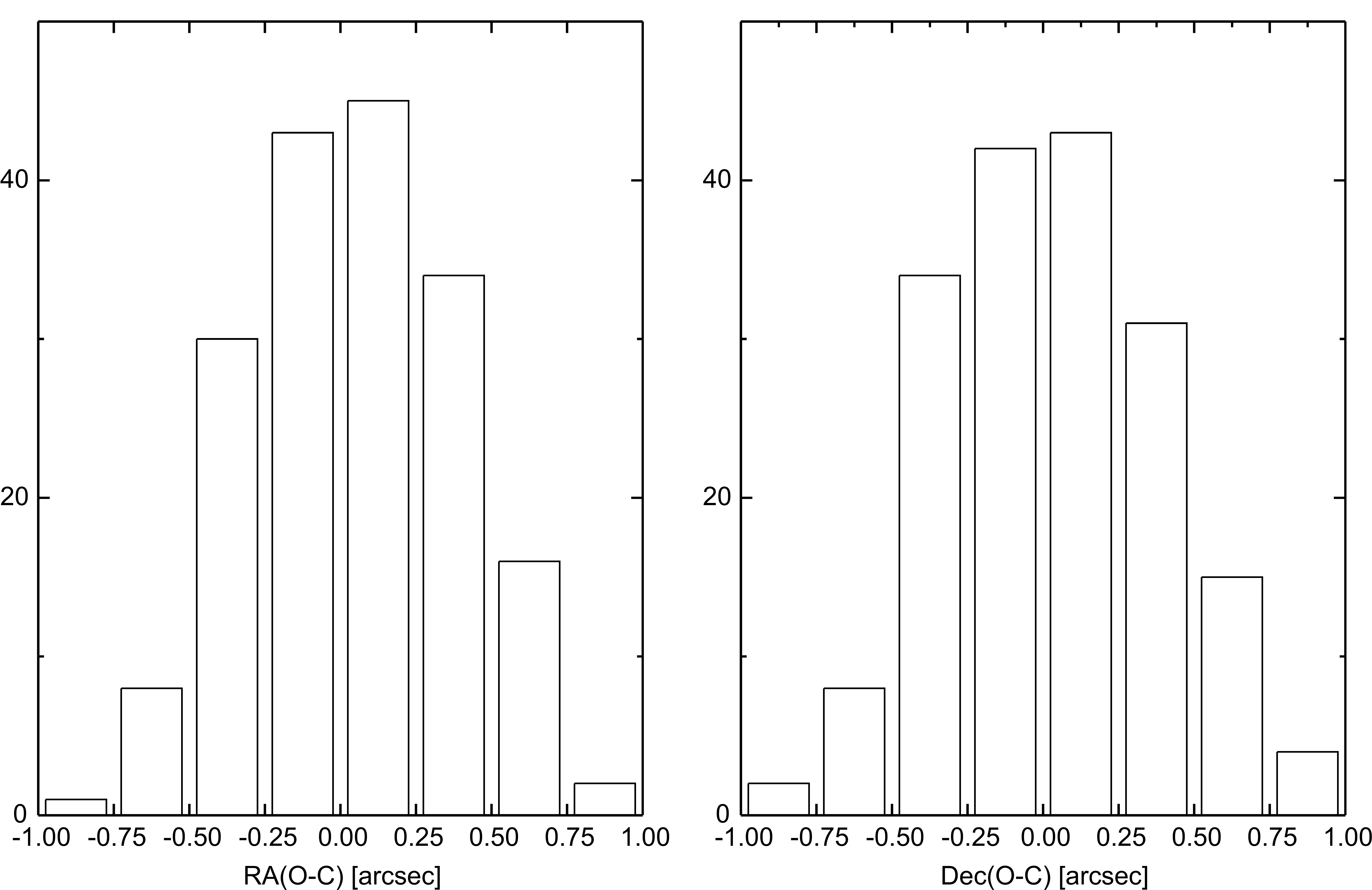}}
\caption{Histograms of the residuals of the derived best-fitting orbital solution of comet 2I/Borisov.}\label{residuals}
\end{figure*}

\end{document}